# A Starter Kit for Diversity-Oriented Communities for Undergraduates

## Chapter 1: Near-Peer Mentorship Programs

**Sponsored by The Access Network, January 2025**

Compiled and written by Emily J. Griffith[1] and Gloria Lee[2], Joel C. Corbo[3] , Gabriela Huckabee[4], Hannah Inés Shamloo[5], Gina Quan[6], Noah Charles[7], Brianne Gutmann[6], Gabrielle Jones-Hall[7], Mayisha Zeb Nakib[9], Benjamin Pollard[10,11], Marisa Romanelli[9], Devyn Shafer[12], Megan Marshall Smith[13], Chandra Turpen[14]

## Author Affiliations

[1] Center for Astrophysics and Space Astronomy, Department of Astrophysical and Planetary Sciences, University of Colorado, Boulder, CO 80309, USA
[2] Department of Natural Sciences, Scripps and Pitzer Colleges, Claremont, CA 91711, USA
[3] Ethnography and Evaluation Research, University of Colorado Boulder, Boulder, CO 80309, USA
[4] Department of Physics, University of California Santa Cruz, Santa Cruz, CA, 95064, USA
[5] Department of Geological Sciences, Central Washington University, Ellensburg, WA 98926, USA
[6] Department of Physics and Astronomy, San José State University, San Jose, CA 95192, USA
[7] Reed College, Department of Physics, Portland, OR 97202, USA
[8] Amazon Fulfillment Technologies & Robotics, North Reading, MA 01864, USA
[9] Department of Physics, University of Illinois Urbana-Champaign, Urbana, IL 61801, USA
[10] Department of Physics, Worcester Polytechnic Institute, Worcester, MA 01609, USA
[11] Department of Physics, University of Auckland, Auckland 1010, NZ
[12] Columbus City Schools, Columbus, OH 43215, USA
[13] Department of Physics, Hamilton College, Clinton, NY 13323, USA
[14] Department of Physics, University of Maryland, College Park, MD 20742, USA

## A short history of this document

This mentoring resource is a collaborative project years in the making. It began as a workbook entitled "Designing a near-peer mentoring program for science majors", which was written by Hannah Shamloo and Anna Zaniewski for an Access Network winter workshop on mentoring that ASU Sundial hosted in 2017. In 2019, Gabriela Huckabee of ASU Sundial, with guidance from Anna Zaniewski, Gina Quan, and Joel Corbo, compiled the best practices from all sites within the Access Network into a comprehensive "starter kit". The mentoring workbook was incorporated as a chapter of this document, which is entitled "A Starter Kit for Diversity-Oriented Communities for Undergraduates" and has been shared internally within the Access Network. In 2021, Access Network Core Organizers Gloria Lee, Noah Charles, and Joel Corbo initiated efforts to shepherd the starter kit to completion, editing content and style in order to publicly disseminate this information. In 2024, Core Organizers Emily Griffiths and Gloria Lee completed the first phase of this public dissemination effort, focusing on the mentoring chapter of the starter kit. They streamlined and updated content, as well as designed supplemental worksheets and activities, with the hopes that student leaders at universities around the country will feel supported as they establish their own near-peer mentoring programs.

# Table of Contents

# I. The Importance of Mentorship

Mentorship provides a number of important benefits for those trying to navigate science[1]. For undergraduate students, these benefits include improved academic performance[2], increased social integration[3], stronger retention rates[4], improving students' beliefs in their ability to succeed[5], and developing a sense of belonging[6] and a more robust personal and professional network[7]. Mentoring programs can also be an important tool for creating more diverse, equitable, and inclusive STEM departments[8]. This is because while informal mentoring tends to be readily accessible to students coming from more well-resourced backgrounds (for example, in the form of "hidden curriculum" or social networks), structured mentoring communities can intentionally provide information and foster relationships across all student backgrounds.

The EP3 Guide to Advising and Mentoring Students[9] put together by APS and AAPT does an excellent job outlining effective practices for mentoring in the physical sciences. The guide introduces skills that individual faculty members should practice to proactively advise and mentor their students, as well as provides information on how to set up formal structures to support mentoring within a department. The guide also emphasizes the importance of reflection, suggesting methods for assessing the outcomes of a given mentoring approach. The studies backing these effective practices are organized in a resources section for further reference.

While there is a wealth of knowledge for faculty advising on academics and research, fewer resources exist for establishing near-peer mentoring programs, which connect more experienced students with newer students to guide them through their college experience. The inter-student relationships and communities that these near-peer mentoring programs foster, however, can play a vital role in helping students navigate science. Beyond helping students develop a sense of belonging in their respective fields, these student-led programs also cultivate student agency and buy-in. These programs promote student ownership of learning environments by providing students with opportunities to design, rather than just experience, more equitable and accessible spaces in the physical sciences, which can feed back into greater ownership of their learning and further enhance the benefits described above.

This mentoring resource is a guide to establishing and running near-peer mentorship programs. It is based on the working knowledge and best practices developed by the Access Network, a collection of nine student-led communities at universities across the country working towards a vision of a more diverse, equitable, inclusive, and accessible STEM environment. Many of these communities, also referred to as sites, include a near-peer mentoring program that is developed to best support their local context. The format of these programs vary, ranging from structured classes with peer mentoring groups to student clubs supporting 1-on-1 relationships. To further support program participants as both students and as whole people, sites often run additional events such as lecture series, workshops, and social activities guided by students voicing community needs. Through this process, student leaders have generated and honed best practices for all aspects of running their sites.

This guide is an attempt to synthesize those efforts, offering practical advice for student leaders setting up near-peer mentorship programs in their own departments. It has been written through the lens of *undergraduate* near-peer mentorship programs, although our framework could easily be extended to other demographics (e.g. high schoolers, graduate students, etc.). Our experience is with STEM mentorship specifically, though these practices can extend to any discipline. We hope to address questions such as: which mentoring format will work best for your local context? What are effective ways to recruit, onboard, and check-in with participants? How do you ensure mentoring relationships are proactive and equitable? and how do you keep up engagement with program alumni and ensure program sustainability?

Many of us know from first-hand experience the invaluable support, perspective, and social ties a good mentoring relationship can provide. Those experiences may be part of your own motivation for starting a mentoring program and for wanting it to be as equitable and sustainable as possible. They are also what led us to write this guide. We hope that it will be useful to you as you build your program.

**Discussion Activity: Think-Pair-Share**

Think of a mentor, advisor, coach or teacher who made an impact on you. What aspect of that relationship was important to you? In general, why is mentorship important?

# II. Designing a Mentorship Program

When launching or restructuring a mentorship program, there are many topics to consider before advertising the program and recruiting participants. To design a mentorship program from the ground up, you must consider important factors such as who will be involved, what will be accomplished, and how the program will be structured. In this section, we will help you identify the who, what, when, where, and why for your program by:

1. Identifying your target audience
2. Setting program goals
3. Selecting a mentorship format
4. Defining mentoring group dynamics

We strongly suggest that your leadership team discuss each topic before proposing your program to department leadership and applying for funding.

## Target Audience

The goals and shape of your mentorship program will depend upon the community you are serving. When determining your target audience, consider who you will recruit as mentees as well as mentors. This may be defined by factors such as your time and resources, the identities and experiences of your leadership team, and the needs of your department or university.

### Mentees

What is the community that you want to serve? Some possible options include:

- Undergraduate students of a particular year
- Undergraduate students of a particular discipline/major
- Transfer students
- Non-traditional students (e.g. part time)
- International students
- Historically excluded demographics (e.g. racial minorities, gender minorities, disabled students)

Consider how your program will fit in with existing mentorship efforts within your department or university, such as a Women in Physics group. How can your program complement or expand upon existing groups without competing for mentees or mentors?

After initially defining the demographics of your target mentee community, estimate the number of students who meet your criteria. If this number is small, consider broadening the scope of your program. If this number is very large (especially compared to your number of potential mentors), consider if you will narrow your community, broaden your mentor pool, or implement an application process. Oftentimes, undergraduate academic advisors or the university admissions department can provide demographic data to help you estimate your community size. Be aware that the provided demographic information may be minimal and may not include information on all identities you wish to serve.

### Mentors

Who is best suited to mentor the community that you want to serve? Some possible options include:

- Graduate students
- Senior undergraduate students
- Past mentees
- Students of a particular discipline/major
- Students with similar demographics to the mentee community
- Students with knowledge of the path through your university's undergraduate STEM major
- Students who have done research

When considering your target mentor and mentee communities, it is important to acknowledge that everyone's identity is multifaceted, with axes of privilege and axes of oppression. How can mentors use their unique lived experiences to help and guide the mentees? How can they use their privileges to support and advocate for the mentees?

As suggested above, we encourage you to estimate the size of your potential mentor pool. Consider the time commitment of mentorship vs. the time that your target mentor community has available for service work.

## Goals of Mentorship

Strong, well-articulated goals will drive the design of your mentorship program. Goals should be set in collaboration with the target mentees. The Access Network advocates for student-led programming—the support that you are offering should serve the actual needs of the mentees, not their perceived needs from someone in a position of power. We suggest revisiting the program goals annually to evaluate if they have been met and to add/remove goals as the program evolves.

When setting goals for your mentorship program, consider:

- **What are the assets and strengths of your mentee community?** Examples include unique experiences, availability of specialized resources, strong home community, etc.
- **What are the challenges faced by your mentee community?** Examples include busy schedules, working multiple jobs and/or caretaking, lack of network, etc.
- **What are the needs of your mentee community?** Examples include networking opportunities, research experiences, scholarships, professional development guidance, etc.
- **What support can your mentors provide?** Examples include research mentorship, discussing their path, introducing mentees to their network, knowledge of undergraduate curriculum, etc.

**Discussion Activity: Write Goals for Your Programs**

Your program should set goals for all participants, including the mentees, the mentors, and the community as a whole. Write a few goals for each group.

By the end of the program, mentees will:

- 
- 
- 

By the end of the program mentors will:

- 
- 
- 

By the end of the program all participants will:

- 
- 
-

# Mentorship Program Format

Mentorship programs can take on many forms. Depending upon your communities goals, size, funding, and level of organizational support, one format may suit better than others. Here we briefly describe four ways to structure your mentorship program and list some of the pros and cons of each. We rate each format from 1 to 5 on the time commitment to organize and sustain as well as the cost, with 1 being low and 5 being high.

## Unstructured Mentor Groups

Time commitment: ⚫⚫◯◯◯
Cost: $ ▢ ▢ ▢ ▢

An unstructured mentoring community offers a space for mentorship to occur with formal mentor-mentee pairing but without a set schedule or agenda. This may be similar to a university club, but without the oversight or funding of a registered organization. A mentoring community can take on the shape that the organizers and participants desire. Unstructured mentorship programs are typically supplemented with organized events or activities, such as a welcome dinner where participants can meet and socialize with each other.

**Pros:** The mentoring relationship is student led and can be continued as long as the mentee and mentor desire. Mentor pairs can meet when and where they prefer and do not have set discussion topics. Beyond the initial recruitment and pairing, there is little week-to-week organizing for the leadership team.

**Cons:** Unless regular community events are held, program participants will have little interactions with members outside of their immediate mentorship group. Leadership has little supervision on how often mentoring pairs meet and what they discuss. This puts the onus on the students to organize meetings and can lead to mentoring relationships dissolving.

## University Club/Student Organization

Time commitment: ⚫⚫⚫◯◯
Cost: $ ▢ ▢ ▢ ▢

A university club offers mentorship programs the flexibility to set their own schedule and the opportunity to develop a large, widespread community. While there may be some oversight from the university itself, the day-to-day activities and structure of the program will be set by the organizers. Beginning an official university club requires the leadership team to submit an application to a student governing body. This application will likely require a purpose statement and proposed plan of events or activities that your club will host.

**Pros:** A club can serve a larger number of people than a classroom or more structured mentorship format could accommodate. Mentor-mentee groups could be set up within the organization and then meet in or outside of the club structure. A club has freedom to meet whenever is convenient for its participants, and does not punish anyone for dropping involvement partway through the semester. University clubs are inexpensive to start, and often are provided funding (or can apply for funding) to host events and purchase materials.

**Cons:** While the lack of regulation and structure allows for flexibility, it requires commitment from participants to become and stay involved. Organizers must be vigilant about ensuring that the club meets and is active. University clubs may require specific roles, such as president, vice-president, and treasurer that may not be cohesive with the less hierarchical structure that Access advocates for.

## Semester Class

Time commitment: ⚫ ⚫ ⚫ ⚫ ◯
Cost: $ $ $ ▫ ▫

A semester-long class can provide structure for a mentorship program and offer space for the entire mentoring community to regularly meet. A classroom format allows for both unstructured time for mentoring groups to meet and provides a space to have group discussions about topics including DEI, professional development skills, and university/departmental resources. Running a semester-long class requires a team of content developers and teachers, including a graduate student teaching assistant (TA) and often an undergraduate learning assistant (LA). Consider who will serve as your TA and/or LAs and how you will recruit for these roles annually. University and department paperwork is usually required to list the class for credit.

**Pros:** Class credit may incentivize undergraduate students to participate. The regularly scheduled class time will ease mentor-mentee meetings, as students do not have to organize amongst themselves. The classroom environment will maintain a wider mentoring community, allowing for mentee-mentee and mentor-mentor interactions. It is also easier to bring in outside resources such as guest speakers in a class format.

**Cons:** Classes are a large time commitment, requiring the development of lesson plans, in-class hours, and office hours. Official approval from the university may be required one or more semesters before your class can be listed with the registrar. Beginning a new TA line may also require time, paperwork, and outside funding. If undergraduate students take the class for credit, you will be required to assign them a grade, so students' academic record could be damaged if they receive low scores.

## Summer Early Start Program

Time commitment: ⚫ ⚫ ⚫ ⚫ ⚫
Cost: $ $ $ $ $

Summer early start programs target incoming first year undergraduate students and introduce students to college, a certain department or topic, and living on their own before the academic year begins. Early start programs are an effective way to create lasting communities among participants[10]. While a future chapter of the starter kit will provide more details on designing and sustaining a summer early start program, formal and informal mentorship can be a key component.

**Pros:** Early start programs can help students adjust to college life and overcome apprehensions about their STEM major. Because students spend 6-12 hours a day with their program peers, students are able to quickly form friendships with their new classmates. Early start programs allow for informal mentorship from program facilitators and could include a formal mentorship component.

**Cons:** Summer early start programs are expensive (> $15,000) as you must consider expenses such as housing, meals, field trips, and stipends for students and facilitators. They also require substantial time commitments from a team of organizers, often including university administrators from multiple offices (e.g. housing, dining services, facilities management). In addition to a larger logistical lift, summer programs offer a specific context for different relationships to form. If your site decides to transition from a summer program to a mentoring program over the semester, there may be additional challenges in keeping up relationships when the structure provided by the summer program is removed. This by itself is not a con, rather another aspect to consider when designing your program.

**Note:** This chapter of the Access Starter Kit does not include information on how to start or run a semester class or pre-arrival program, but we hope to cover this in a future chapter.

**Discussion Activity: Think-Pair-Share**

Think of your community, their identities, their needs, their obligations, and their constraints. What platform or combination of platforms will serve your community best? Write how each would or would not support your program vision and goals.

Unstructured Mentor Groups:

University Club:

Semester Class:

Summer Early Start Program:

# One-on-One vs Group Mentoring

Peer mentorship, where mentors and mentees are close in age or academic progress, can occur in both one-on-one and group mentorship settings. Depending upon the goals of your program and the ratio of mentors to mentees, one-on-one or small group mentoring may be best for you. Below we summarize the pros and cons of each group structure.

## One-on-One Mentoring

In one-on-one mentoring, mentees are directly paired with a mentor. We discuss how to pair mentors and mentees according to their preferences in Section Suggestions for Matching Mentors and Mentees.

**Pros:** One-on-one mentoring provides mentors and mentees with a space to form a strong bond and have personal communication. The mentee has individualized feedback and support. Scheduling meeting times is easier than with a small group. The mentor is not at risk of being stretched thin by multiple mentee's needs.

**Cons:** The success of a mentor-mentee pair is dependent upon the rapport between individuals. While some pairs may work well with one-on-one meetings, others may struggle to maintain conversation. If one individual cannot make the meetings or leaves the program, the pairing cannot continue. Paired mentorship requires an equal recruitment of mentors and mentees and may limit the size of your program.

## Group Mentoring

In group mentoring, multiple mentees are paired with one mentor or a small group of mentors. A more senior undergraduate and graduate student can co-mentor a group to provide a range of experiences and perspectives. Mentoring groups can be defined by mentor and mentee preferences, research area, or other shared interests. Group mentorship may work best when paired with a semester class or another mentorship platform with a regular meeting time.

**Pros:** Mentees have access to many different perspectives and academic paths. Mentees can bond with other mentees in the group, and mentors can support each other in mentorship. With multiple mentors in a group, undergraduates who have completed the program as a mentee can be recruited as mentors while still having support from other senior group members. This form of mentoring group structure allows for fewer mentors than mentees, allowing a program to serve more students.

**Cons:** In a larger group setting, mentees may not develop a deep connection with a particular mentor. If group mentorship is used with an unstructured platform, it may be difficult for a large group to find a time when everyone is available to meet. If a single mentor has many mentees, they may become overwhelmed if all students need individualized support.

You do not necessarily have to make a choice between one-on-one and group mentorship: it's possible to pair some mentors and mentees one-on-one and place others in groups. This may, however, make matching more complicated as you will need to pair mentors and mentees not just based on their individual compatibility, but also based on the structure they prefer.

**Discussion Activity: Think-Pair-Share**

For your program, which of the pros and cons are the most compelling to determine your mentorship group structure? Are there other benefits or difficulties of mentoring group size that are not listed?

# Worksheet: Program Design

In Section VII: Appendix Resources, we include a Program Design Worksheet. This worksheet is intended to guide prospective mentorship program leadership through the choice of target audience, mentorship group structure, and mentorship platform by considering the resources available and program goals. This worksheet could also be utilized by existing mentorship programs looking to reflect upon their program structure.

# III. Running a Mentorship Program

Below we outline the basics of running a mentorship program, including:

- Recruiting mentors and mentees
- Matching mentors and mentees
- Orienting mentors and mentees
- Checking in with mentors and mentees and collecting feedback
- Running supplemental programming

We will discuss each of these topics in this section and the following sections.

## Logistics Timeline

This is an example logistics timeline for how to run a mentoring program as a university club/student organization, adapted from Illinois GPS. In this format, program leadership takes care of mentor-mentee matching, runs an initial orientation meeting, and plans social and academic events throughout the school year.

**End of Academic Year**

Send out applications for next academic year!

1. Create or update application forms as necessary.
2. Publicize by giving short presentations in classes or student organization meetings, emailing undergrad, grad student, and program listservs, and posting on social media.
3. Ask STEM student organization leaders to publicize to their members.
4. Ask college and department for a list of incoming freshmen and transfer students to email.

Ask for student feedback from current year.

1. Create or update feedback forms as necessary.
2. Email program listserv, and send frequent reminder emails and follow up with individuals to get responses. This is mainly to determine whether students want to continue with the program, and if so whether they want to remain with their current mentors/mentees.
3. Update program listserv by removing students who will not continue with the program.

| **Summer** |
| --- |
| Match mentees with mentors.<br>1. If necessary, sort applications into accept and waitlist lists (create and use guidelines for this process). *Note: our sites tend to prioritize students who have been previously waitlisted for participation in the following semester.*<br>2. Generate mentor and mentee profiles from the applications and organize a matching session.<br>3. Once matches are generated, create a roster and update program listservs.<br><br>Inform mentees and mentors about their new matches.<br>1. If necessary, inform waitlisted students of their waitlist status.<br>2. Email new program members, welcoming them to the program.<br>3. Email mentor-mentee matches. It may be useful to provide mentors and mentees with some intro information about each other (gathered from their applications) if they have not met before. |
| **Beginning of Academic Year** |
| Plan mentor and mentee orientations.<br>1. Ideally, orientation will take place right before a welcome social for the entire program at the beginning of the school year.<br>2. You can give a presentation on expectations and best practices for mentors and mentees, overview how the program will run, and invite some previous mentors and mentees to talk about their experiences.<br>3. The social afterwards then functions as an easy way for new mentors and mentees to meet each other for the first time, or former mentors and mentees to catch up. Pro-tip: provide name tags 🙂<br><br>Plan social and academic activities.<br>1. Hold regular meetings to plan and organize events. At the first meeting, you may wish to create a schedule of activities based on student interest, and assign a point person for planning each event. Some example activities include research lab tours, a fall-themed social, journal clubs, etc. It may also help to partner with other STEM student organizations to run specific events.<br>2. If necessary, create online forms or polls to determine when members are free during the week to participate in activities, as well as gauge interest in different activities. |

3. Reserve rooms as necessary and request funding from your department or university.
4. Advertise events to program listserv and on social media.

# Recruiting Mentors and Mentees

In this section, we describe various methods for recruiting student participants as well as organizers. Keep in mind that some methods will be more effective with reaching certain groups. For example, high schoolers never pick up the phone, so call their parents instead.

## Email

The easiest tool for recruitment is to send emails to listservs. A recruitment email typically includes:

- [ ] A concise pitch about your program, including who you serve, the benefits of joining the program, who you are looking to recruit, and any expected time commitments
- [ ] Information on how to apply (link to application form, emails of point people, etc) and any relevant deadlines
- [ ] Links to social media or emails for more information or questions

Utilize school- or department-wide listservs to reach a wide audience. Ask your department's administrative office for different group listservs and email etiquette! Your college admissions office, freshmen and transfer-student advisors, or your home department may be able to provide a list of incoming students or send out information on behalf of your site. Examples of current Access sites that do this are IMPRESS at Rochester Institute of Technology, which has the College of Liberal Arts and Sciences send invitation emails to all deaf/hard-of-hearing and first generation students, and CU-Prime at University of Colorado Boulder, which has the Physics Office Manager and freshmen advisors send out CU-Prime information emails to freshmen who have declared an interest in physics.

If there are other student groups that your target population overlaps with (for example, you want to recruit physics students and your institution has a Society for Physics Students), then you can work with those groups to advertise through their listservs as well. Emails tend to be the most wide-reaching and general recruitment technique, allowing you to recruit undergrads, grad students, and faculty to take on whatever role is necessary for your site.

## Snail Mail

In addition to emailing information about your program to your target population, sending a physical letter of invitation in the mail can be a powerful tool to recruit students. The letter should include a welcoming introduction, a summary of the program, and information on how to apply or learn more. We suggest printing letters on university letterhead and sending with university envelopes to look official–there is likely an office on campus that can assist. If you've developed a brochure or flyer for the program, this should be included as well. Academic advisors and/or the admissions office can provide contact information for undergraduate students.

## Phone Banking

Included with contact information for students will be a phone number for them or their parents. Calling and/or texting is another effective and personal method to reach your target students. Write a short script to read if someone answers the phone, making sure to ask if they have questions or would like more information. Have another script ready if the call goes to voicemail. Following up the call with a text that includes links to your website and program application easily connects students with more information.

## Referrals

Referrals are a high impact way of reaching individual students. Ask faculty, advisors, and leaders of clubs for students who they think may be a good candidate for participation in the site, and reach out to those individuals. This approach is especially useful for finding mentors in a particular field, or identifying students who may benefit from mentoring but don't typically respond to email blasts. Referrals are also often one of the only ways to reach incoming freshmen (via academic advisors). If your site is aimed at first year students, be sure to cultivate a good relationship with the academic advisors who will be working with them. These are the same people that will be helpful in sending mass emails to incoming students! Additionally, if there are student organizations present that have similar goals to your site or are related in some way, you could ask their leaders to join your site as well or recommend their own students.

## Word of Mouth and Networking

Having site members plug your program to people in their networks and classes is a great way to bring trusted people into your community. Typically, if students are brought in by an existing member, they may already have vested interest in your mission and pre-existing ties to other community members. This method is effective for bringing in new students and also a really strong method for recruiting site organizers and teachers if your program runs a class.

## In-Class Presentations

Ask your faculty allies if you can give a short pitch in their classes. It can be more impactful if you can bring a student who went through your site's programs to talk about the benefits. Target the community that you are trying to serve. For example, if your target community is freshmen, advertise in introductory-level classes.

## New Student Orientations

Send representatives to new student orientations to pitch your site and collect contact information. Most incoming freshmen and transfer students will attend an in-person orientation and will be looking for ways to get ahead!

## General Advertising

Go back to the basics - put up some flyers! Make sure to follow campus regulations when advertising, but flyers left on public posting boards, department announcement boards, and at the front desks of resident halls and advising offices can be helpful in spreading information. Post in school-wide social media groups or start a new group for your site. You can use this to post pictures, highlight student success stories, and announce events as well!

## Tips and Tricks

When you advertise your site, you should include a link to a contact email or an application so that interested people can stay connected to your community. An application through Google forms is usually the best approach, as it allows you to collect emails, names, contact info, and any kind of demographic or academic/personality information that you may need for evaluating your site or matching mentees with compatible mentors. The GPS site leadership team uses Google forms to collect information on mentees and mentors so that they can make the best matches possible. Template application forms are included in the Matching Mentors and Mentees section below. Paper application forms left in strategic locations can also be helpful!

One of the most effective recruitment tools is in-person information sessions. Provide free food as an incentive to come and hear about your site, and ask participants from previous years to come and help talk about how the site impacted them. Ask everyone who shows up to introduce themselves and why they came. Have a short presentation that covers: site goals, time commitment, incentives for participation, and expected activities. Cool loot is always a big plus (undergrads love stickers). Use many (or all!) of the methods suggested above, you're likely to catch different students through different channels.

A word on recruiting leaders and organizers specifically: the best place to look for leaders is among the students in your site. Access has a focus on student leadership and advocacy, and students that are active participants in the site can be mentored by current leaders to become more independent and take on more responsibility. Giving a student a bite-sized project to start with, such as helping plan a community event, will give them a sense of agency and deeper connection to the site and community.

# Matching Mentors and Mentees

There are many ways to match mentors with mentees. Some examples include group "speed-matching" events, using a "stable marriage" algorithm, or manually matching people. Regardless of the method, you usually want to start with an application that allows you to collect some basic information about your mentors and mentees. The information included in an application is useful for both selecting program members and generating mutually beneficial matches.

The example application questions included below are intended to prompt discussion of educational or personal backgrounds and reasons for joining the program. Depending on your specific program and method for matching mentors and mentees, you may choose a subset of questions to use or generate your own. A subset of the answers can then be used to generate short bios of each mentee and mentor when introducing them to each other.

## Template Mentee Application

This is a list of questions that may be helpful to ask potential mentees, adapted from applications developed by Illinois GPS.

| **Basic Information** |
| --- |
| ☐ Name<br>☐ University email address<br>☐ Year<br>☐ Transfer Student?<br>☐ Major<br>☐ Do either of the following apply?<br>    ☐ Neither of your parents have a college degree<br>    ☐ You are a Pell Grant recipient |
| **Short Answer Questions** |
| ☐ Why are you applying to the program, and what do you hope to get out of this experience? |
| ☐ What's the coolest thing you've learned about the natural world recently?<br>☐ Think about a time when you felt a barrier or challenge to your success. Tell us how you overcame it.<br>☐ What has your experience been in STEM classes so far?<br>☐ (Optional) Tell us anything else you'd like about yourself. |
| **Personality Questions** |
| ☐ What do you like to do in your spare time?<br>☐ What do you want to talk to a mentor about? (ranked list)<br>○ Advice on classes (What classes should I take?)<br>○ Advice on career path (How do I get into research or grad school?)<br>○ What is research like?<br>○ Philosophical discussions of science<br>○ Life advice<br>○ Let's just hang out<br>☐ How often would you like to meet during the school year? |
| **Demographics** |
| ☐ Gender identity<br>☐ Where are you from?<br>☐ First language<br>☐ What ethnic group(s) do you most identify with?<br>☐ Religious affiliation (if any)<br>☐ Do you identify as part of the LGBTQ+ community?<br>☐ We asked you a bunch of questions about your identity in this section. Are there any aspects of your identity that you would like us to consider for matching you with a mentor? Are there any that you would specifically like us to not consider?<br>☐ Dietary restrictions (for event planning)<br>☐ How did you hear about us? |

## Template Mentor Application

This is a list of questions that may be helpful to ask potential mentors, adapted from applications developed by Illinois GPS. Some questions may apply specifically to graduate student mentors.

| **Basic Information** |
|---|
| ☐ Name<br>☐ University email address<br>☐ Year<br>☐ Research Area |

| Short Answer Questions |
| --- |
| ☐ Tell us what your undergrad experience was (or is) like. (Include whatever is relevant to your mentoring perspective. Did you always know what you wanted to major in? Did you ever participate in a mentoring program?)<br>☐ (Optional) What kind of guidance do you feel most (or least) comfortable providing? |
| **Personality Questions** |
| ☐ What do you like to do in your spare time?<br>☐ What do you want to talk to a mentee about? (ranked list)<br>○ Advice on classes (What classes should I take?)<br>○ Advice on career path (How do I get into research or grad school?)<br>○ What is research like?<br>○ Philosophical discussions of science<br>○ Life advice<br>○ Let's just hang out<br>☐ How often would you like to meet during the school year? |
| **Demographics** |
| ☐ Gender identity<br>☐ Where are you from?<br>☐ First language<br>☐ What ethnic group(s) do you most identify with?<br>☐ Religious affiliation<br>☐ Do you identify as part of the LGBTQ+ community?<br>☐ We asked you a bunch of questions about your identity in this section. Are there any aspects of your identity that you would like us to consider for matching you with a mentee? Are there any that you would specifically like us to not consider?<br>☐ Dietary restrictions (for event planning)<br>☐ How did you hear about us? |

## Guidelines for Selecting Mentors and Mentees

If your program has a maximum capacity and you have more mentor and/or mentee application than room in the program, you may have to admit and reject participants from the program. For both mentors and mentees, we suggest evaluating how well the participants application aligns with your program goals and the community you wish to serve. If you must reject mentees, consider providing the students with other opportunities for support at your university and/or providing them priority in a future year of programing. Consider if/how your program could expand to serve the greater needs of the community. Regardless of application number, mentor applications should be evaluated to ensure that mentors' goals align with your organization's mission, especially on equity and inclusion. If you feel a mentor is unprepared or unwilling to serve your mentee community or fully participate, do not admit them to the program.

Once you have selected your mentees and mentors, create a roster and listserv for your program. It may be helpful to create separate listservs for mentees and mentors as well if you anticipate holding separate events (i.e. orientation events, town halls for feedback) or asking for different information (i.e. research panelists) from them.

## Suggestions for Matching Mentors and Mentees

There are many potential methods to pair mentors and mentees. Below we list a few potential options. Keep in mind that pairing for one-on-one vs. group mentorship will require different matching algorithms.

**Speed Matching:** In this matching method, mentors write short bios that are sent out to all potential mentees. Mentors and mentees then mingle at a welcome event, rotating to talk to different match candidates they are interested in for set amounts of time. After the event, mentees fill out a form to rank how interested they are in specific mentors (eg: from Not At All Interested to Extremely Interested on a 5 point scale). Program leadership meets to match as many people as possible with the highest interest possible, taking into account other considerations such as class schedules and desired discussion topics to break potential ties.

**Stable Marriage Algorithm:** A stable marriage algorithm (or Gale-Shapley algorithm) is a mathematical method of matching two sets of elements given their preferences or rankings. To apply this algorithm to mentee-mentor pairs, each group must rank order their preferences for the opposite group. With a small number of mentors and mentees, a "speed networking" event can be held where each mentee is able to talk with each mentor. Both groups then rank their preferences. With a large number of participants, each mentee will only be able to meet with a smaller subset of mentors. An initial ranking based on mentor bios could be used to determine pairs for the speed networking event or a mixer could be held for mentors and mentees to mingle. Once rankings are in hand, the stable marriage algorithm can be implemented through Mathematica, Python, or your favorite coding language.

**Manual Pairing:** As the name suggests, in manual pairing the leadership team takes the mentee and mentor applications and/or pairing preferences and pairs mentorship groups by hand. This method of pairing can take into account aspects such as matching research interests between the mentee and mentor, and may be better for setting up mentorship groups.

# Setting Clear Expectations

To set clear expectations for mentors and mentees, program leaders may wish to have participants sign a "Mentoring Agreement" which acts as a contract acknowledging the role of each participant in a mentoring relationship. Specific expectations should be discussed during orientation or training sessions, so that both parties can recognize and agree upon their respective roles over the course of the program. It is very important to specify mentor reporting requirements (for example, many grad students are mandatory reporters under Title IX). Included below are example expectations for mentors and mentees. You may consider adapting these into a mentoring contract which all members of a mentoring relationship read, agree upon, and sign. More details on expectations and responsibilities for mentors and mentees are discussed in Section IV of this mentoring resource.

<table>
<tr><th>A mentor is:</th><th>A mentor is not:</th></tr>
<tr><td>
<ul>
<li>a guide with experience and knowledge who is committed to the mutual growth of the mentor and mentee.</li>
<li>a caring facilitator who helps their mentee make use of resources and increase their network.</li>
<li>a trusted ally and advocate who works on behalf of their mentee's best interests.</li>
<li>someone who sets high expectations and has a high level of belief in the capabilities of their mentee.</li>
</ul>
</td><td>
<ul>
<li>a (surrogate) parent.</li>
<li>a professional counselor or therapist.</li>
<li>a tutor.</li>
<li>a research mentor.</li>
<li>a romantic partner.</li>
<li>judgmental.</li>
<li>given to gossip.</li>
</ul>
</td></tr>
<tr><td colspan="2">As a mentor, you agree to:</td></tr>
<tr><td colspan="2">
<ul>
<li>listen with the intention of understanding your mentee's perspective.</li>
<li>commit to meeting at agreed-upon times and communicate if conflicts arise.</li>
<li>contact site leaders if you have a concern or run into difficulties.</li>
<li>understand your limits and refer your mentee to external resources where appropriate.</li>
<li>respect the confidences of your mentee.</li>
<li>not discriminate against your mentee based on religion, national origin, ethnic heritage, race, sexual orientation, gender identity, or disability.</li>
<li>abide by the university student code of conduct.</li>
<li>follow university required mandated reporting and ensure your mentee is aware of what you must report.</li>
</ul>
</td></tr>
</table>

| As a mentee, you agree to: |
| --- |
| • proactively communicate your goals and progress with your mentor.<br>• commit to meeting at agreed-upon times and communicate if conflicts arise.<br>• contact site leaders if you have a concern or run into difficulties.<br>• understand your mentor's limits.<br>• respect the confidences of your mentor.<br>• not discriminate against your mentor based on religion, national origin, ethnic heritage, race, sexual orientation, gender identity, or disability.<br>• abide by the university code of conduct. |

**Discussion Activity: Think-Pair-Share**

How would you write a mentor-mentee contract for your program? From the items listed, are there any that don't apply to your program? Are there any other points that you would add?

# Building a Mentoring Relationship

## Proactive Communication

The foundation for a beneficial mentoring relationship is proactive communication. Begin by deciding how you want to communicate and meet. For example, you may want to check in every couple weeks over text and figure out times to meet in person. Or you may schedule recurring video calls once a month.  Many people have found it helpful to always put a day in their calendar for when to meet next. If you choose to use a mentoring contract, you can specifically include an agreement on how often to meet/communicate and by which methods. One of the most common challenges in a mentoring relationship is staying in contact as life gets busy and plans change. Communicate through these busy seasons - it is always okay to let the other person know you are busy, and it is always okay to reach back out after a period of non-interaction. Don't let shame or guilt get in the way of clearly communicating.

Once you've established your communication norms, proactive communication will help you get the most out of a mentoring relationship. For mentees, this means clearly communicating your goals at the moment, your successes and struggles reaching them, and questions that you have. For mentors, this means practicing active listening, asking specific questions to guide your mentee in their thinking process, and connecting them with resources or other people that may provide useful information in their journey.

## Conversation Prompts

Good mentoring conversations help mentees reflect on their current reality and generate options for moving towards their goals. On a personal level, they can also help a mentee feel seen, understood, or not alone in their experiences. Below are some example prompts for starting an insightful and meaningful conversation, drawn from the STEM Equals Project[11]. If a mentee seems interested in discussing a topic further, listen actively and ask specific, clarifying follow-up questions. Be okay with silence as well! It may mean that your conversation partner is considering a new perspective or putting a collection of thoughts together.

**Life**

- What have you recently been working on/learning about that you find exciting?
- What has been on your plate recently?

**Professional Development**

- What goals are you working towards at the moment? What steps have you taken so far/want to try taking to reach your goals?
- What kind of support do you need to reach your goals?

**Academics**

- How did you do on your ______ project/exam/assignment?
- Have you tried a new study strategy? How is that working?

**Career**

- What are some career options you're excited about pursuing?
- What is one idea you have for exploring this option further?

## Activities

Planning an activity together is an easy way to keep a date in the calendar and generate experiences to discuss. Activities can be simple ways to hold a conversation, such as meeting at a coffee shop or going for a walk. They may be ways to learn together, such as attending a scientific talk or agreeing to read a book on a topic you both find interesting. Activities like these have the added benefits of built-in discussion topics. If mentors and mentees share similar interests, it may be fun to share those activities as well, such as playing a sport, attending a musical performance, visiting a local attraction, etc.

# How to Check In

Some mentor/mentee relationships may thrive from the get-go, but others may need some additional support or guidance throughout the program. Checking in with mentors and mentees is beneficial for student leaders to assess how each relationship is progressing and identify any potential problems. The format of these check-ins might depend upon the mentorship format you choose.

For mentorship formats with less facetime between program participants (e.g., unstructured mentorship groups and university club/student organization), we recommend circulating a monthly check-in form to the mentors and mentees individually. A template for this form is shown below, and asks about the frequency of meetings, the methods of communication/meet-up, the topics that have been discussed, as well as if there are any concerns or questions for the program organizers. A reminder should be sent to mentors and mentees every month to fill out the form, and the responses should be read and addressed in a timely manner. If a pair consistently stops checking in, program organizers should reach out to the mentor and mentee individually. While unstructured mentoring relationships occasionally dissolve, you want to ensure that the mentee's needs are being met.

| **Frequency and method of meeting** |
|---|
| • How many times this month have you talked with your mentor/mentee?<br>• What are your main methods of communication? (e.g. email, text, in person)<br>• Where do you meet? (e.g. coffee shop, for walks, at colloquium) |
| **Discussion topics** |
| • In our recent meetings, my mentor/mentee and I discuss (select all that apply)<br>○ Academics/study habits<br>○ Career paths<br>○ Topics in Physics/STEM<br>○ Time management/work-life balance<br>○ Professional development/skill and tips<br>○ Diversity, equity, and inclusion topics<br>○ Shared interests<br>○ Personal struggles<br>○ Personal successes<br>• Is there anything you would like the program organizers to know? Do you have any concerns with the mentorship program or your mentorship relationship? |

A similar style of check-in form can also be used for more structured platforms (e.g., semester class and summer early start). Such check-in forms could be circulated virtually, assigned as homework, or completed in class. In a more structured setting, program organizers are able to keep track of attendance and will have a better sense of each mentorship group's successes and needs. Again, we recommend reaching out to mentors and mentees directly if attendance or participation drops off. In a structured setting, open discussions of the mentoring relationships could also occur, encouraging group reflection on what is going well and what changes could be made for the future.

# IV. Mentor and Mentee Training

Mentorship and menteeship training is vital to prepare both the mentors and mentees for a successful mentoring relationship. Mentors need to be prepared to support students in a variety of circumstances, and know where to turn when they alone cannot address a mentee's needs. Mentees need to learn what types of support their mentors can provide and how to communicate their thoughts and concerns to their mentor. Both need to know the expectations of the program, of each other, and how to address conflict, should one arise. In this section, we outline the important aspects to both mentor and mentee training. We provide training lesson plans and associated activities in the Appendix Resources.

## Training Facilitation

Mentor and mentee training serves two main purposes: (1) to introduce participants to the program and their responsibilities and (2) to encourage reflection on what it means to be a mentor or mentee. We strongly encourage mentorship programs to hold a training session where mentors and mentees can actively engage with the content, learn from their peers, and be vulnerable in sharing their own experiences, successes, and failures. Creating an environment where everyone feels comfortable to show up as their full selves, to learn, and to grow is key to successful training. Consider the following ideas when determining who should facilitate and attend your training session, when your training session should be held, and where your training should take place.

### Who?

**Facilitator:** The training facilitator should be familiar with the mentorship program and should ideally be someone with mentoring experience. The facilitator should be prepared to lead complex discussions about mentorship and identity in STEM, to uplift minority voices/perspectives, and to challenge participants to confront their own biases and weaknesses. If using the materials provided in this guide, the facilitator should thoroughly review all content, complete the mentoring exercises themselves, and adapt the lesson plans to be program specific.

**Participants:** As the title of the section suggests, we encourage mentorship programs to run both mentor and mentee training as it is critical to prepare both parties for their involvement in a mentoring relationship. We suggest holding a joint mentor/mentee training session so that mentors are aware of mentee's expectations and vice versa, and then splitting the mentors and mentees into separate groups for reflection activities.

## When?

Mentor and mentee training should be hosted before the start of the mentorship program. If tied to an academic course, we suggest hosting mentor-mentee training during the first class period. Holding the training session early allows both mentors and mentees to be prepared for their new mentoring relationship. Along with this starter kit, we include lesson plans for a two hour training session. This could be hosted as one two-hour session or two one-hour sessions, depending upon your programs' needs. Beyond the initial training, some mentorship programs hold additional training sessions or workshops throughout the year to continue expanding the mentor's tool kit.

## Where?

**In-Person:** In-person mentor-mentee training gives participants an opportunity to meet each other and talk in a more casual manner. Consider hosting the training in a room with space for small group discussion. If hosting mentor and mentee training concurrently, you may want two adjoining rooms to separate mentors and mentees for some portions of the training. We suggest holding the training within the host department so that mentees can be introduced to department spaces. Keep in mind that sensitive topics may arise in group discussions, so consideration should be made for a closed door space where non-participants cannot intrude or overhear.

**Virtual:** While we encourage in-person training when possible, virtual meetings allow for participants to join from any location, and without the health risks of close-contact interactions. This format may be particularly useful if training is held over the summer when mentors and mentees are not on campus. To encourage engagement, it is important to set virtual meeting norms, such as having cameras on (when available), muting microphones when not speaking, using the raise hand and yes/no features, and refraining from multitasking on another screen. We suggest the use of breakout rooms to facilitate small group discussions.

# Key Components of Mentor Training

## Setting Expectations

A mentoring relationship is a two-way street. The key to a rewarding mentoring relationship is that all parties are proactive in communicating their expectations. Mentoring relationships also change and grow, as both mentee and mentor move through different phases in their academic careers and personal lives. It is good practice to check in periodically to see whether your expectations for the mentoring relationship are being met or need to evolve.

## Expectations and Responsibilities of Mentors

**Be proactive:** Mentors should take the lead on the organizational aspects of the mentoring relationship such as setting meetings times, reporting back to program leaders, and having topics prepared to discuss. Mentors should also be proactive in modeling the type of mentoring relationship we wish to foster. This includes modeling vulnerability and honesty, leading discussions on the "unspoken" aspects of academia, and fostering a supportive, compassionate, and positive space[12,13]. To encourage open discussions, mentors can propose discussion topics and help mentees identify what they feel comfortable talking about, then follow up with specific questions about that topic.

**Listen:** A critical piece of effective mentorship is clear communication and active listening. Active listening requires attentiveness to what the mentee is sharing, understanding their thoughts, reflecting on what is being said, and retaining the information that was shared. Mentors should avoid judging or interrupting their mentee and should respect their mentee's experiences and perspective, even if it is at odds with their own. We encourage active listening through mirroring, where the listener mirrors what they have heard back to the speaker, validates the emotion or experience that was shared, and empathizes with the speaker's experience.

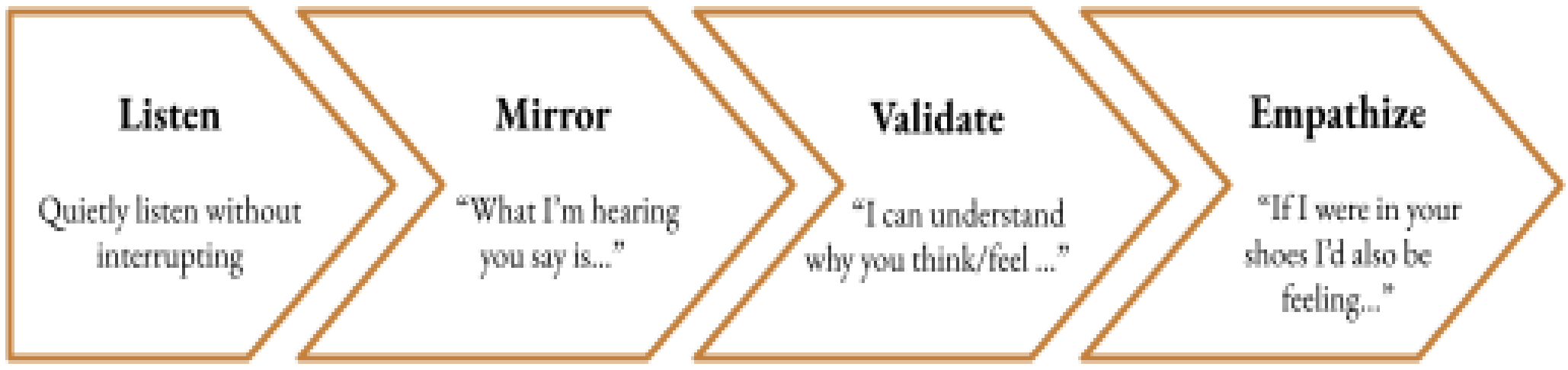

**Reflect:** Mentors should take time before and after meeting with their mentees to reflect on their role as a mentor, their experiences as an earlier career scientist, how their identities shape their own personal experiences, and how they can best support their mentee. Mentors should also be encouraged to reflect on their position of power in the mentoring relationship, considering how the power imbalance might impact their mentee's willingness to be vulnerable, as well as how to use their position of power to help their mentee. This reflection can be explicitly encouraged through check-in forms, as discussed in How to check in, and the mentor training activities provided in Section VII: Description of Appendix Resources.

Beyond compassionate and empathetic mentoring, there are likely program-specific responsibilities that your mentors need to know. Some specific responsibilities of a mentor might include:

- Meet with their mentee on a bi-weekly basis
- Submit a check-in to program organizers after every meeting
- Attend a colloquium, group meeting, job fair, etc. with their mentee
- Discuss summer research opportunities and provide feedback on written materials
- Create and guide mentee through a small research project

In addition to setting clear expectations for mentors, it is also important to explicitly state the roles that mentor does not fill. For example, a mentor is NOT: a tutor, a parent, a therapist, or a taxi. Mentors should be prepared to help students access additional resources when needed rather than feel like they are their mentee's lone support system. Mentor training can reassure mentors that it is okay to admit that they are not capable of meeting all of their mentee's needs, and that other avenues and resources are there to help mentees access the support they need. In addition, if mentors are employed by the university, they are likely mandatory reporters for sexual harassment and/or sexual assault. It is important that mentors and mentees know each other's reporting duties at the start of the mentoring relationship.

## Expectations and Responsibilities of Mentees

**Be proactive:** Mentees are inherently proactive in their own mentorship as they have identified a need for near-peer support and signed up for your mentorship program. This should be acknowledged and appreciated! Mentees are expected to continue being proactively involved in their mentorship by showing up to mentoring meetings, making the mentoring relationship and mentorship program homework a priority, and investing time in their professional development work that may be associated with the program.

**Communicate:** It is critical for a mentorship program to foster an environment where mentees are comfortable talking openly and honestly, communicating their needs, goals, successes, and struggles. Just as mentors are expected to listen, mentees are expected to discuss their academic interests, their experiences in STEM, and their needs from the mentoring relationship. These concepts can be difficult to communicate, though, as mentees may not know what they need or even what mentorship can offer them. Part of the facilitator's and mentors' role is to help mentees learn how to communicate these ideas. The mentee-specific activities in the mentor training lesson plans provide an example of how to start this conversation.

**Reflect:** Mentees should take time before the start of the mentorship program and periodically throughout their mentorship to reflect on their goals, both personal and academic, their mentoring relationship, and their own experiences within academia. At first, this reflection should be guided by mentors or facilitators to encourage mentees to consider many axes of their experiences. The Access Identity Activity in Section VII: Appendix Resources is one possible resource to guide this reflection. When mentees reflect on their struggles or negative experiences within academia, it is important to highlight how toxic academic systems contribute to such feelings, normalize imposter thoughts, and separate academic success from personal worth.

Other program specific responsibilities of a mentee might include:

- Meet with their mentor on a bi-weekly basis
- Submit a check-in to program organizers after every meeting
- Attend a professional development event
- Write a CV, resume, and/or personal statement for a summer research program application
- Complete weekly course assignments

## Program Norms

Beyond mentor and mentee expectations, it is important to set norms that all program participants and facilitators agree to follow in large group and one-on-one interactions. Norms establish how participants will engage in discussion with each other in a respectful and constructive manner. We encourage the program facilitators to suggest a set of norms at the start of mentor/mentee training and lead a group discussion on what norms to adopt. Norms should be revisited through the program to ensure that they are effectively promoting equitable discussions and supported by all participants.

Within Access, we adopt the following norms at our annual Assembly. They are designed to uplift all voices in large group discussions amongst diverse participants.

- Be civil and respectful when you have disagreements.
- Respect other's decisions to participate, or not participate. It's OK to excuse yourself.
- Faculty, graduate students, and others in positions of power are encouraged to let others speak first during open discussions. Moderators should prioritize diverse voices.
- Build in pauses after questions to allow everyone time to think before responding.
- Use consensus cards to aid in conversation. Use finger snaps to indicate agreement.
    - The purpose of the consensus cards is to assist in consensus-based decision making. They are used to make sure everyone has a voice in the conversations. The different colors of the consensus cards indicate:
    - Green: Yes, I agree. Jazz Hands. Yes vote. I support you / what you said.
    - Yellow: I have a question / comment. My vote is hesitant.

- Red: Wait! I disagree. Please get back on topic. That violates our norms. No vote.
- Abide by the "don't yuck my yum" principle–don't disparage viewpoints that may not align with your own

## Providing Resources

While near-peer mentorship programs offer valuable personal and, occasionally, academic support to their participants, mentors should not be expected to be a student's full support system. Mentors are not academic advisors, tutors, mental health professionals, or crisis managers—but they may find that their mentees need such forms of support over their academic career. Common struggles faced by undergraduate students include:

- Failing a course
- Needing to drop or add classes from their schedule
- Finding affordable housing and meal plans
- Navigating roommate or classmate conflicts
- Dealing with family emergencies
- Struggling with depression and anxiety
- Accessing academic accommodations
- Balancing a work schedule
- Applying for scholarships

and much more. When mentees are vulnerable and share their very personal struggles with their mentors, this should be met with compassion, validation, and empathy. We expect mentors to support their mentees through such challenges by providing a safe space to talk openly about their successes and struggles, but not by solving these problems for them. Instead, we want to encourage mentors to help their mentees build a larger support network and access other resources that are available within the university or larger community.

To do this, mentors (and mentees) need to know what resources are available. Incoming undergraduate students likely will not know about all the support structures within the university and most graduate student mentors will not be familiar with the undergraduate program or college resources. As a part of mentor training, we encourage mentorship programs to provide an overview of departmental and university resources, as well as a summary of undergraduate major pathways and academic advising plans. This information should be presented at the mentor training event and kept in an accessible place where mentors and mentees can access it when needed.

## Departmental and University Resources

Most undergraduate-serving institutions have academic, research, health, financial, and other resources available to their students. However, students may not know of the resources' existence or how to access them. Students may also have a fear of asking for help or feel as if they have to navigate college on their own. Creating a centralized list of resources and encouraging mentors to share these resources with their mentees can lower the barrier to access and grow a student's support network.

Prior to mentor training, we suggest that the mentorship program leaders research the local resources and create an accessible list of programs or offices that can provide support to the students served by the mentorship program. We encourage making this list inclusive of the needs for minoritized populations and those that are often left behind–make sure to list institutional resources as well as student groups! In Section VII: Description of Appendix Resources, we include a template student resources brochure that could be personalized to your university and mentorship program.

**Tip:** Reach out to the new student orientation groups! They may already have a general list of resources compiled!

Categories and examples of resources to consider:

- **Academic:** Tutoring services, math learning center, writing center, libraries
- **Career:** Career counseling, job/internship fairs, resume workshops
- **Department Specific:** Colloquium/talk series, summer research programs
- **Accommodations:** Disability services
- **Mental Health:** University counseling services, community counseling services, crisis hotlines
- **Physical Health:** Student health center, gyms, wellness center
- **Advocacy:** Office of diversity and inclusion, multicultural center, legal services, safety/escort service, military and veteran services, international affairs office, Title IX office, student organizations for marginalized groups such as LGBTQ+, disabled, neurodiverse, Black, Latin and Hispanic, Indigenous tribal members, Middle Eastern, Asian, multiracial and/or international students
- **Financial:** bursar, financial aid office, emergency funds/grants, food pantry
- **Safety:** safe walk programs, safe ride shares, community knowledge regarding safety issues on or near campus

**Discussion Activity: Think-Pair-Share**

What resources were critical to your success as an undergraduate? Make a list of five resources available at your university and how to access them.

1.
2.
3.
4.
5.

## Academic Advising

Over the course of the mentoring relationship, mentees will likely discuss their current course work as well as their future academic plans with their mentor. While mentors should not serve as an academic advisor, it is important that they know the basics of completing a STEM undergraduate major and the diversity of paths through it. For example, a mentee starting in college algebra may be nervous about completing the prerequisite math courses needed to begin Physics I. If the mentor is aware of a four year plan to complete the physics major, beginning in college algebra, they can provide the mentee with guidance and reassurance that a physics major is still possible.

Talking with departmental academic advisors is critical to accessing this knowledge. Check with your department to see if there is a staff member or university office that can help. Academic advisors can often provide resources including a course catalog with course numbers, class descriptions, and pre-requisites as well as example four or five year plans for the completion of the major with different starting points, semesters abroad, and/or additional majors or minors. This information may be available on university and department websites as well.

**Tip:** Invite the academic advisor to your training! They likely have valuable knowledge to share with the mentors and mentees. Mentees will likely need to talk with their academic advisor at multiple points in their undergraduate career, so providing an additional space for students and advisors to meet will be beneficial.

## National Resources

Beyond resources available at your university, there are national resources, scholarships, and organizations that can offer career, academic, social, and financial support to students. Many of these resources are centered around a specific identity, and may help students from underrepresented backgrounds find a deeper community. Consider making a list of national resources that mentees in your program may be interested in. Look for organizations or scholarships specific to your discipline and mentee demographics.

# Discussing Conflict Resolution Strategies

## Preparing for Potential Conflict

Despite the best efforts of site leaders and program members, occasional conflicts will arise in mentoring relationships. Communication may break down, misaligned expectations may lead to frustration or misunderstanding, or other personal issues and circumstances may get in the way of a healthy mentoring relationship. Below we outline some techniques that individuals may find helpful in resolving conflict within their mentoring relationship. It is important to discuss conflict resolution with both mentors and mentees during training so that both parties are prepared if conflict arises. We also suggest a structure that programs can use to help their members resolve conflict in an organized way, such as an ombuds team (described below).

## Navigating Conflicts

Mentors and mentees should be prepared to respectfully navigate conflicts that may arise within their relationship and should also know who to turn to if the conflict cannot be resolved. We recommend covering the following strategies in mentor/mentee training to navigate interpersonal conflict:

- Deal with the conflict promptly. Don't let it sit and fester. The other person may have no idea they are hurting you and would appreciate the opportunity to correct their actions as soon as possible.
- Start by assuming best intentions, and that the conflict arose from a misunderstanding or misalignment of expectations.
- Be honest in communicating what happened and how that made you feel. If your intention is to continue the relationship, consider that in the statements you make as well.
- Use "I" statements.
- Actively listen to each other and use the mirroring technique.

- If you are in the wrong, own your mistake and apologize with sincerity. Say "I'm sorry I made you feel that way", not "I'm sorry you felt that way."
- Offer concrete actions you can take to remedy the situation or make sure this mistake does not occur in the future.

If a conflict is not resolved, the aggressor is not responsive to the tactics outlined above, or more serious conflicts arise (e.g., harassment, insensitive/hurtful comments, belittlement, non-responsiveness), the mentor or mentee should bring the conflict to the leadership or ombuds team, depending on their comfort level and needs.

## Importance of an Ombuds Team

Near-peer mentoring programs are great because oftentimes program leaders are close to program participants and can design programs that closely hew to what participants need. However, one disadvantage is that if conflict arises, oftentimes those very same program leaders are personally connected to those having conflicts, making it difficult to resolve conflicts in an impartial, impersonal manner. One solution to this problem is to designate an ombuds team[14] through which conflict can be resolved in a neutral and informal manner.

An ombuds-person is a site member who others can go to to informally discuss issues or concerns they may have within the program. The ombuds acts as a neutral third-party observer and helps guide mentors and mentees through possible solutions, all while maintaining confidentiality of those reporting the issues. Further, if an ombuds-person notices systemic issues being reported, they may bring these issues up to program leadership for resolution, also maintaining confidentiality. Sites may want to designate several members as ombuds-people and form an ombuds team. This way, if an ombuds-person does play another role in the organization that may compromise their neutrality during a particular conflict, there are other ombuds that site members can go to for help.

| An Ombuds-Person Does: | An Ombuds-Person Does Not: |
|---|---|
| • serve in an informal capacity.<br>• listen to understand the perspective of those bringing up concerns and issues.<br>• maintain neutrality and confidentiality.<br>• help individuals develop possible solutions and evaluate different approaches to resolve their conflict. | • formally report to the program, department, university, etc.<br>• take sides.<br>• keep records.<br>• make decisions for any individual or issue any corrective measures.<br>• serve in any other role in the organization that may compromise their neutrality. |
| • encourage members to directly address their concerns, pointing them to effective techniques and/or other resources. | |

# Best Practices for Inclusive Mentorship

To set both mentors and mentees up for success, we recommend sharing mentorship best practices. These tips will guide mentors to create a safe and empathetic environment for the mentees, and will orient the mentees on how to ask for help from the mentors. We suggest providing both groups with these lists, in addition to the further reading section below. To reinforce these practices during mentorship training, complete the role-playing exercise found in the Appendix of mentorship training materials.

## Mentor Best Practices

A mentor's role is to foster a relationship with the mentees, creating a space to provide academic/career advice, discuss struggles and successes, and be honest about the realities of being a student in STEM. To be an effective mentor requires empathy, vulnerability, understanding, and reflection. As many mentorship programs target students from historically underrepresented backgrounds on axes of race, ethnicity, gender, sexuality, disability, income, parental education, citizenship status, and more, it is critical to recognize how a mentee's identity may shape their experience in STEM, in college, and in society. Many people of minoritized identities have been systemically discriminated against and systemic barriers exist in STEM for students from these backgrounds. Mentors must recognize that different identities afford different privileges and face different means of oppression. Individuals may experience an intersection of these identities and experience both privileges and external biases. Being mindful of your own identity, your mentee's identity, and the interaction of your identities in the context of society will be useful in developing cultural awareness and increasing the chance of a trusting mentoring relationship. A further reading list on how identity shapes STEM experiences is included in Section VI.

To foster such a safe space and be an effective mentor to all students, we suggest the following strategies:

- **Do not dismiss or diminish experiences that mentees share.** Listen to learn, educate yourself on systemic challenges to STEM students from underrepresented backgrounds, challenge your own biases, and take action to confront discrimination.

Do not generalize mentee's identities or struggles. Every student should be treated as an individual, and their unique intersectional identity should be recognized.

- **Create a space where students feel comfortable talking about their specific struggles and feel comfortable bringing their whole selves to the table.** Culture shock and imposter syndrome are common. Acknowledging identity-specific challenges can validate students' experiences and make them feel less alone. Be mindful of power structures within the relationship and how this might impact what the mentee chooses to share or discuss.
- **Encourage students to join, find, or create community.** Being the "first" or "one of a few" is an isolating experience, and students may have had to learn to be independent/self-reliant early on. Social support networks play a key role in increased well-being, success, and retention[15].
- **Emphasize the breadth of career options available to STEM students in addition to academia.** Financial, familial, disability accommodations, and citizenship considerations may be important factors for underrepresented students in choosing their careers. Acknowledge that scientists are human beings first and have other aspects to their life outside of intellectual passions and career development that are also important and valid.
- **Help students find resources and unpack some of the "hidden curriculum" in academia.** Connect them to professional organizations designed for different identities in STEM, support them in applying for research programs, and encourage them to present at conferences. Help students find financial aid or apply to paid research and internship opportunities (remember to be mindful of citizenship requirements). Help students grow their network by introducing them to colleagues.
- **Make referrals when needed**. If an issue comes up that is beyond the scope of your mentoring relationship, refer the student to resources that are appropriate for the context. Consider supporting the mentee in using resources by offering to help them make an appointment or find the appropriate building.
- **Demonstrate faith in the abilities of the students to overcome barriers and achieve success.** Be a cheerleader in the good and bad moments. Provide feedback in constructive ways without judgment and without diminishing the student's abilities.
- **Do not assume students' identities.** Instead, create a space where students feel comfortable discussing their individuality with you. Remember that many aspects of identity, such as disability, class, sexuality, gender, and ethnicity, cannot be seen. It is the mentee's choice to disclose or not disclose any aspect of their identity.
- **Discuss the underrepresentation of minority scientists** and help students find role models. If there are no faculty at your university that your mentee can identify with, help them identify scientists in other institutions or within national groups (e.g. Black in Physics) that could be future mentors or role models.

- Be mindful of money–**don't ask students to go on expensive excursions** or meet for lunch/coffee at an expensive location. Don't make assumptions about your mentee's financial situation. Ask program organizers if the mentorship program could cover costs if needed.
- **Educate yourself!** As mentors, it is critical that we continue to learn about the experiences of others in STEM and what we as individuals can do to be changemakers locally. Acknowledge your own shortcomings and seek out trainings, book clubs, town halls, etc. to learn more.

Beyond this general advice, mentors should be mindful of considerations for specific identities:

- Mentors should be aware that if they are in the majority race while mentee is in the minority race, there may be a structural barrier to trust.
- Recognize and confront racism whenever you can. Participate in bystander, microaggression, and anti-racism trainings if available.
- Use a student's correct name and pronunciation. Know an appropriate time and place to ask for a student's pronouns, such as in a one-on-one or small group setting, rather than in front of an entire class.
- Do not assume documentation status. Familiarize yourself with local laws about undocumented students and eligibility for in-state tuition. Many research programs require citizenship status, make sure to recommend opportunities that are open to everyone
- Seek out additional training opportunities, such as DREAMzone or UndocuAlly, to learn more about being undocumented in the US.
- Respect mentees' accommodations and needs. If you are unsure of how to support a mentee with specific accommodations, consult with the campus disability office.
- Recognize that Deaf culture (with a capital "D") may exist at your university. Educate yourself on the meaning of being deaf, non-verbal, etc., and ask students what terms they prefer to describe themselves.

## Mentee Best Practices

A mentee's role is to communicate with the mentor, verbalizing where they need personal or academic support. This practice requires the mentee to reflect and be proactive, which may be difficult at the start of a new mentoring relationship. These skills require practice, but with the right mentor and a supportive environment, mentees may feel more comfortable discussing vulnerable topics. As a mentee, it is common to not know what to ask for, or what types of support your mentor can provide. To get the most out of your menteeship, we suggest the following strategies:

- **Ask for help.** Your mentor is there to support you through your successes and struggles. If peer mentoring relationships are new to you, it might feel like you are asking too much of your mentor. This is a normal feeling, but remind yourself that it is your mentor's role to help you, and that they want to provide you with that support. Some examples of things you can ask your mentor for help with are:
    - Finding resources, such as finding a tutor, accessing mental health resources, or identifying the right office on campus to help.
    - Providing feedback on application materials, such as Research Experiences for Undergraduates (REU) applications.
    - Discussing instances of discrimination in the classroom or elsewhere. Your mentor can also support you in escalating concerns to faculty or department heads.
    - Validating feelings of imposter syndrome or stereotype threat.
    - Brainstorming ways to build community or enact change in your department
- **Take time to reflect on your experiences.** This is perhaps the most difficult and most important part of mentorship. By reflecting on your own experiences, you may find it easier to describe your emotions and to discuss vulnerable topics with your mentor. This will help your mentor to provide appropriate guidance and support.
- **Try new things and meet new people.** While it can be daunting to attend a STEM colloquium, a job fair, or your mentor's research group meeting, these are great networking opportunities and will expose you to potential career opportunities.
- **Ask questions.** Whether attending a colloquium, discussing a new topic in your STEM class, or considering applying to graduate school, you likely have questions that you don't know how to find the answers to. These questions, personal or scientific, can be great conversation topics during your mentorship meetings. If your mentor doesn't know the answer, you can find it together!
- **Take notes.** Some of the guidance your mentor may be able to provide might not be immediately applicable to you, such as advice on applying to graduate school if you are a first year undergraduate. This information is still useful to have, so remember to jot down some notes for future you.
- **Stay in touch with your mentor.** While the official mentoring relationship may have a fixed length, your mentor will likely continue to be a supportive figure throughout your academic career. Reach out to your mentor if you have questions about class registration, research opportunities, or job applications in the future. They may even be able to write you a letter of recommendation!

# V. Post Program Engagement

Effective mentoring relationships can last long past the period of formal engagement with a mentorship program, and setting up structures that support continued interactions with program alumni can help maintain the sustainability and longevity of your program.

## Keep an Alumni List

Maintaining a list of mentees and mentors who have been in your program will make it easier to maintain contact with students who have participated in the program. This practice is helpful for tracking retention and the long term careers/successes of your alumni.

## Host Annual Social Events

Hosting annual events is a great way to bring together past and present program members. Graduation celebrations are a popular option–students can reflect on the community that shaped their path to this large academic milestone, celebrate each others' successes to date, and inspire each other with their future endeavors. It is also a great way to publicize where alumni are headed next, be it industry, graduate school, etc., and to maintain those connections should they be helpful to future program members.

## Recruit Past Mentees as Mentors or Program Leaders

Having experienced firsthand the impact of a positive mentoring relationship, your mentees may want to return as mentors. Inviting alumni who are now upperclassmen is a great way to recruit mentors and to keep past mentees engaged in mentorship. Alumni may also want to get involved in program leadership, event planning, or other aspects of your organization. Their input is critical, as the program should be designed to serve their needs!

## Invite Alumni to Participate in Mentorship Events

While some alumni may not have time to mentor or take on a leadership position, they may still want to contribute to the mentorship program. You can keep these alumni engaged by asking them to participate in one-off events such as panels and workshops to share their perspectives as they move along their educational and career journeys.

# VI. Description of Appendix Resources

These resources are available online and can be downloaded from the following link: https://accessnetwork.org/starter-kit/

## Program Development

**Program Design Worksheet:** A worksheet to guide program organizers through the design of (or reflection on) a near-peer mentoring program. This worksheet asks you to consider your target audience, the pros and cons of mentorship group structure, the pros and cons of different mentorship platforms, and the overarching goals of your program.

## Mentor-Mentee Training

**Mentor-Mentee Training Lesson Plans:** Lesson plans for a 2.5 hour training session, describing the training objectives, materials needed, facilitator preparation, and outline for the training session, including how to use the other resources listed here.

**Mentor-Mentee Training Slideshow Template:** A slide presentation to be used for mentor and mentee training. The slides should be adapted to your specific needs, but contain many stand alone subsections on topics such as mentor/mentee expectations, active listening, identity and privilege, and setting goals.

**Mentor Identity Worksheet:** A worksheet designed for mentors' reflection on their own identity, the ways in which various identities are dominant or non-dominant, and how identity impacts the ways in which we interact with and experience the world.

**Mentee Road Map Worksheet:** A worksheet designed for mentees' reflection on their goals, their existing support structure, and the support/connections that they want to build over the course of the program.

**Mentor Case Study Cards:** Three pairs of mentor/mentee cards describing the mentor/mentee's perspective on an example scenario, designed for mentors to practice navigating challenging situations. The cards contain the scenario on one side and discussion questions on the other.

**Mentee Case Study Cards:** Three pairs of mentor/mentee cards describing the mentor/mentee's perspective on an example scenario, designed for mentees to practice communicating concerns to their mentor. The cards contain the scenario on one side and discussion questions on the other.

**Resource Pamphlet Template:** A template tri-fold pamphlet to be filled with academic, career, mental health, physical health, financial, department specific, and advocacy resources.

# Acknowledgements

Special thanks to Sarah Monk and Robert Strausbaugh for style editing. We thank all past and current Network Fellows, Assembly Fellows, Core Organizers, and Access Assembly attendees for joining discussions and sharing their experiences with near-peer mentorship. Thank you to site leaders from The Compass Project, IMPRESS, Chi-Sci Scholars, Sundial, Equity Constellations, CU-Prime, GPS, North Star, and Polaris for sharing information about and resources from local mentorship programs.

This material is based upon work supported by the National Science Foundation under Grant Nos. 1506129, 1506190, 1506235, 1806516, 1806709, 1806668, 1806566, 1806585, 2011895, 2011972, 2011953, 2011877, 2011892, 2011780, 2309307, 2309308, 2309309, 2309310, and 2309311. Any opinions, findings, and conclusions or recommendations expressed in this material are those of the author(s) and do not necessarily reflect the views of the National Science Foundation.

E.J.G. is supported by an NSF Astronomy and Astrophysics Postdoctoral Fellowship under award AST-2202135.

# References and Further Reading

The following resources go further in depth on how different social identities may shape student experiences in STEM fields and college.